\documentstyle[aps,floats,amssymb,graphics]{revtex}

\begin{document}
\draft
\title{XAFS spectroscopy. II. Statistical evaluations in the fitting problems}
\author {K.~V. Klementev}
\address{Moscow State Engineering Physics Institute,
115409 Kashrskoe sh. 31, Moscow, Russia\\ e-mail: klmn@htsc.mephi.ru}
\date{\today}
\maketitle

\begin{abstract}
The problem of error  analysis is addressed in stages beginning with the case of
uncorrelated parameters and proceeding to the Bayesian problem that takes into
account all possible correlations when a great deal of prior information about
the accessible parametr space is available. The formulas for the standard
deviations and deviations with arbitrary confidence levels are derived.
Underestimation of the errors of XAFS-function extraction is shown to be
a source of unjustified optimistic errors of fitting parameters. The
applications of statistical $\chi^2$- and $F$-tests to the fitting problems are
also discussed.
\end{abstract}
\pacs{61.10.Ht}

\section{Introduction}
In the Open Letter to the XAFS Community \cite{Young1} Young and Dent,
the leaders of the UK XAFS User Group, expressed their concern over the
persistence of lingering common opinion that XAFS is a ``sporting technique''
and it is possible to obtain the ``answer you want''. Some way out they see in
a special attention to the publishing XAFS data (first of all, to XAFS spectra)
and have formulated several recommendations for editors and referees.
Undoubtedly, in the matter of extraction of the real, not invented, information
from XAFS experiments the quality of spectra is of great importance. We see
here another problem as well. Not having some necessary elements of XAFS
analysis (some values and the procedures for their determination), one has a
quite natural desire to turn those values to advantage. Principally we mean the
inability of the standard methods to find the errors of the atomic-like
background $\mu_0$. Traditionally, the noise is assigned to these errors.
However, as was shown in Ref.~\cite{Ie}, the noise is {\em essentially} lower
than the errors of the $\mu_0$ construction. Below, we will show that the
underestimation of the errors of XAFS-function extraction is a source of the
unreasonable optimistic errors of fitting parameters.

Practically all known programs for XAFS modeling \cite{Catalog} in some way
calculate confidence limits of fitting parameters. However, since there is no
standardized technique for that and since most published XAFS works do
not contain any mention of methods for estimation of the errors of fitting
parameters, the accuracy of the XAFS results remains to be field for trickery.

In the present article we derive the expressions for the errors of fitting
parameters under different assumptions on the degree of their correlation.
Besides, the prior information about parameters is possible to take into
account in the framework of Bayesian approach. Moreover one can find the
most probable weight of the prior information relative to the experimental
information.

We also discuss the grounds and usage of the statistical tests. The special
attention was focused on that where and how one can embellish the results and
artificially facilitate the statistical tests to be passed.

All methods and tests described in the paper are realized in the program
{\sc viper} \cite{VIPER}.

\section{Errors in determination of fitting parameters}
Let for the experimental curve ${\bf d}$ defined on the mesh $x_1,\ldots,x_M$
there exists a model ${\bf m}$ that depends on $N$ parameters ${\bf p}$. In
XAFS fitting problems as ${\bf d}$ may serve both $\chi(k)$ (not weighted by
$k^w$) and $\chi(r)$. The problem is to find the parameter vector
$\hat{\bbox{\rm p}}$ that gives the best coincidence of the experimental and
model curves. Introduce the figure of merit, the $\chi^2$-statistics (do not
confuse with the symbol of XAFS function):
\begin{equation}\label{A1}
\chi^2=\sum_{i=1}^{M}\frac{(d_i-m_i)^2}{\varepsilon_i^2},
\end{equation}
where $\varepsilon_i$ is the error of $d_i$. The variate $\chi^2$ obeys the
$\chi^2$-distribution law with $M-N$ degrees of freedom. Of course, for the
given spectrum ${\bf d}$ and the given model ${\bf m}$ the value of
$\chi^2$ is fully determined; we call it ``variate'' bearing in mind its
possible dispersion under different possible realizations of the noise and
the experimental errors of $d_i$ extraction.

Often a preliminary processing (before fitting) is needed: smoothing,
filtration etc. Naturally, during the pre-processing some part of the
experimental information is lost, and on the variates
$\xi_i=(d_i-m_i)/\varepsilon_i$ additional dependencies are imposed (before,
they were bound solely by the model ${\bf m}$). It is necessary to determine
the number of {\em independent} experimental points $N_{\rm ind}$. For the
commonly used in XAFS spectroscopy Fourier filtering technique the number of
independent points is given by \cite{Stern1}:
\begin{equation}\label{A2}
N_{\rm ind}=2\Delta k\Delta r/\pi+2,
\end{equation}
where $\Delta k=k_{\rm max}-k_{\rm min}$ and $\Delta r=r_{\rm max}-r_{\rm min}$
are the ranges in $k$- ¨ $r$-spaces used for the analysis, and $r_{\rm min}>0$.
If $r_{\rm min}=0$ then
\begin{equation}\label{A3}
N_{\rm ind}=2\Delta k\Delta r/\pi+1.
\end{equation}
Instead of keeping in the sum (\ref{A1}) only  $N_{\rm ind}$ items which
are equidistantly spaced on the grid $x_1,\ldots,x_M$, it is more convenient to
introduce the scale factor $N_{\rm ind}/M$:
\begin{equation}\label{A4}
\chi^2=\frac{N_{\rm ind}}{M}\sum_{i=1}^{M}\frac{(d_i-m_i)^2}{\varepsilon_i^2}.
\end{equation}
Now the variate $\chi^2$ follows the $\chi^2$-distribution with $N_{\rm ind}-N$
degrees of freedom. It can be easily verified that with the use of all available
data ($r_{\rm min}=0$ and $r_{\rm max}=\pi/2dk$) the definition (\ref{A4})
turns into (\ref{A1}).

Let us now derive the expression for the posterior distribution for an
arbitrary fitting parameter $p_j$:
\begin{equation}\label{P1}
P(p_j|{\bf d})=\int\cdots dp_{i\ne j}\cdots P({\bf p}|{\bf d}),
\end{equation}
where $P({\bf p}|{\bf d})$ is the joint probability density function for
all values $\bf p$, and the integration is done over all $p_{i\ne j}$.
According to Bayes theorem,
\begin{equation}\label{P2}
P({\bf p}|{\bf d})=\frac{P({\bf d}|{\bf p})P({\bf p})}{P({\bf d})},
\end{equation}
$P({\bf p})$ being the joint prior probability for all $p_i$, $P({\bf d})$ is
a normalization constant. Assuming that $N_{\rm ind}$ values in ${\bf d}$ are
independent and normally distributed with zero expected values and the standard
deviations $\varepsilon_i$, the probability $P({\bf d}|{\bf t})$, so-called
likelihood function, is given by
\begin{equation}\label{P3}
P({\bf d}|{\bf p})\propto\exp\left(-\chi^2/2\right),
\end{equation}
where $\chi^2$ was defined above by (\ref{A4}). Its expansion in terms of
$\bf p$ near the minimum ($\nabla\!_p\chi^2=0$) which is reached at
${\bf p=\hat p}$ yilds:
\begin{equation}\label{P4}
P({\bf d}|{\bf p})\propto
\exp\Bigl(-\frac{1}{4}({\bf p-\hat p})^T\cdot\tensor{\rm H}\cdot({\bf
p-\hat p})\Bigr)\equiv \Bigl(-\frac{1}{4}\sum_{k,l=1}^{N}\frac{\partial^2\chi^2}
{\partial p_k\partial p_l}\Delta p_k\Delta p_l\Bigr),
\end{equation}
where $\Delta p_k=p_k-\hat p_k$, and the Hessian $\tensor{\rm H}$ components
(the second derivatives) are calculated in the fitting program at the minimum
of $\chi^2$. The sufficient conditions for the minimum are
$\tensor{\rm H}_{kk}>0$ and
$\tensor{\rm H}_{kk}\tensor{\rm H}_{ll}-\tensor{\rm H}_{kl}^2>0$, for any
$k,l$. Hence, the surfaces of constant level of $P({\bf d}|{\bf p})$ are
ellipsoids.

\subsection{Simplest cases}
If one ignores the prior then the posterior probability density function
$P({\bf p}|{\bf d})$ coincides with the likelihood $P({\bf d}|{\bf p})$. Let us
consider here two widely used approaches.

(a) {\em Parameters are perfectly uncorrelated}. In this case the Hessian is
diagonal and
\begin{equation}\label{P4_1}
P(p_j|{\bf d})\propto\exp\Bigl(-\frac{1}{4}\tensor{\rm H}_{jj}\Delta p_j^2\Bigr).
\end{equation}
The standard deviation of $p_j$ is just
\begin{equation}\label{P4_1_1}
\delta^{\rm(a)}p_j=(2/\tensor{\rm H}_{jj})^{1/2}.
\end{equation}

(b) {\em Parameter $p_j$ essentially correlates solely with $p_i$}.
In this case
\begin{eqnarray}\label{P4_2}
P(p_j|{\bf d})&\propto&\int dp_iP(p_ip_j|{\bf d})\propto
\int dp_i\exp\Bigl(-\frac{1}{4}\tensor{\rm H}_{jj}(\Delta
p_j)^2 -\frac{1}{2}\tensor{\rm H}_{ij}\Delta p_j\Delta p_i
-\frac{1}{4}\tensor{\rm H}_{ii}(\Delta p_i)^2\Bigr)\nonumber\\
&\propto&\exp\Bigl(-\frac{1}{4}\left[\tensor{\rm H}_{jj}-
\tensor{\rm H}_{ij}^2/\tensor{\rm H}_{ii}\right](\Delta p_j)^2\Bigr),
\end{eqnarray}
from where one finds $\bar p_j=\hat p_j$ and the mean-square deviation
\begin{eqnarray}\label{P4_3}
\delta^{\rm(b)}p_j=\left(\frac{2\tensor{\rm H}_{ii}}
{\tensor{\rm H}_{jj}\tensor{\rm H}_{ii}-\tensor{\rm H}_{ij}^2}\right)^{1/2}.
\end{eqnarray}
In practice, to find the strongly correlated pairs of parameters, one finds the
pair-correlation coefficients:
\begin{eqnarray}\label{P4_4}
r_{ij}=\frac{\langle\Delta p_i\Delta p_j\rangle-
\langle\Delta p_i\rangle\langle\Delta p_j\rangle}
{\delta(\Delta p_i)\delta(\Delta p_j)}
\end{eqnarray}
taking on the values from -1 to 1. Two parameters are uncorrelated if their
correlation coefficient is close to zero. It is easy to calculate the average
values over the distribution (\ref{P4_2}):
$\langle\Delta p_i^2\rangle=2\tensor{\rm H}_{jj}/{\rm Det}$,
$\langle\Delta p_j^2\rangle=2\tensor{\rm H}_{ii}/{\rm Det}$,
$\langle\Delta p_i\Delta p_j\rangle=-2\tensor{\rm H}_{ij}/{\rm Det}$, where
${\rm Det}=\tensor{\rm H}_{jj}\tensor{\rm H}_{ii}-\tensor{\rm H}_{ij}^2$.
Notice, by the way, that these are the elements of the inverse matrix of
$\tensor{\rm H}/2$. Now the pair-correlation coefficients are given by:
\begin{eqnarray}\label{P4_5}
r_{ij}=-\frac{\tensor{\rm H}_{ij}}{\sqrt{\tensor{\rm H}_{ii}\tensor{\rm H}_{jj}}}.
\end{eqnarray}
Via the correlation coefficient the mean-square deviations, found for the cases
(a) and (b), are simply related:
\begin{eqnarray}\label{P4_6}
\delta^{\rm(a)}p_j=\delta^{\rm(b)}p_j\sqrt{1-r_{ij}^2}.
\end{eqnarray}

\begin{figure}[!t]
\begin{minipage}[c]{0.6\hsize}\includegraphics*{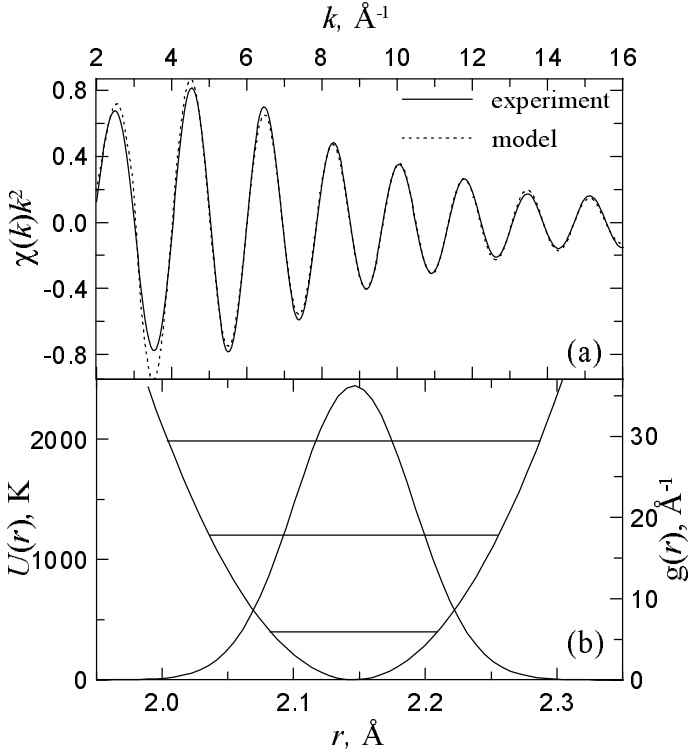}\end{minipage}
\begin{minipage}[c]{0.35\hsize}\caption{Experimental and model filtered
XAFS $\chi(k)\cdot k^2$ (first coordination sphere) for BaPbO$_3$ (a) and
the model potential with corresponding PRDF and energy levels (b).}
\label{PbO}\end{minipage}\end{figure}
Consider an example of the error analysis. For $L_3$ Pb absorption
spectrum\footnote{The spectrum was recorded at 50\,K in transmission mode at
D-21 line (XAS-13) of DCI (LURE,Orsay, France) at positron beam energy 1.85 GeV
and the average current $\sim250$\,mA. Energy step --- 2\,eV, counting time ---
1\,s. Energy resolution of the double-crystal Si [311] monochromator (detuned
to reject 50\% of the incident signal in order to minimize harmonic
contamination) with a 0.4\,mm slit was about 2--3\,eV at 13\,keV.} for
BaPbO$_3$ compound the average error of the XAFS extraction from the measured
absorption was $\varepsilon_i=0.007$. For the filtered over the range
$1.0<r<2.1$\,\r{A} (the signal from the octahedral oxygen environment of lead
atom) XAFS (see Fig.~\ref{PbO}), the model function was calculated as follows.
For one-dimensional the Hamiltonian of the lead-oxygen atomic pair with
potential $U=a/2\cdot(r-r_0)^2$ we found the energy levels and corresponding to
them wave functions. Then, averaging over the Gibbs distribution, the pair
radial distribution function (PRDF) normalized to the coordination number $N$
was found as:
\begin{equation}\label{PRDF}
g(r)=N\sum_n|\Psi_n(r)|^2e^{-E_n/kT}\Bigr/\sum_n e^{-E_n/kT},
\qquad N=\int g(r)\,dr,
\end{equation}
and the XAFS function as:
\begin{equation}\label{gChi}
\chi(k)=\frac{1}{k}
F(k)\!\int\limits_{r_{\rm min}}^{r_{\rm max}}\!\!g(r)\sin[2kr+\phi(k)]/r^2\,dr.
\end{equation}
The phase shift $\phi(k)$ and the scattering amplitude $F(k)$ were calculated
using {\sc feff6} program \cite{FEFF}. By variation of the parameters $r_0$,
$a$, $N$ (where $N$ includes the factor $S_0^2$), and $E_0$, the shift of the
origin for the wave number $k$, one search for the best accordance between the
model and experimental curves. Here for the fitting, the {\sc viper} program
was used which, in particular, calculates the Hessian of $\chi^2$ (defind by
(\ref{A4}) with $N_{\rm ind}=11.8$) at the minimum. The correlation
coefficients are listed in the Table~\ref{CMatrix}.

\begin{table*}[!h]
\center \caption{Pair-correlation coefficients $r_{ij}$ for the example
fitting.} \begin{minipage}{7cm} \begin{tabular}{l@{\quad}*{4}{d@{\quad}}}
&$N$&$a$&$r_0$&$E_0$\\\hline
$N$&1&$-$0.286&0.092&0.041\\
$a$&$-$0.286&1&$-$0.044&0.048\\
$r_0$&0.092&$-$0.044&1&0.727\\
$E_0$&0.041&0.048&0.727&1\\
\end{tabular}\end{minipage}\label{CMatrix}\end{table*}
\begin{table*}[!h]
\center
\caption{Mean values and mean-square deviations of the fitting parameters.
$\delta p$ are the mean-square deviations calculated: for perfectly
uncorrelated parameters (a), trough the maximum pair correlations (b), from the
bayesian technique without prior information (maximum likelihood) (c), from the
posterior probability that considers the most probable contribution of
the prior information. $S_p$ are the sizes of the parameter space accessible
for variation ($\pm$ around the mean value).}
\begin{minipage}{14cm}
\begin{tabular}{l@{\quad}*{6}{d@{\quad}}}
$p$&$\hat p$&$\delta^{\rm(a)}p$&$\delta^{\rm(b)}p$
&$\delta^{\rm(c)}p$&$S_p$&$\delta^{\rm(d)}p$\\\hline
$N$&4.05&0.090&0.094&0.096&$\hat N$&0.070\\
$a$, K/\r{A}$^2$&2.28$\cdot10^5$&4.7$\cdot10^4$&
4.9$\cdot10^4$&4.9$\cdot10^4$&$\hat a$&6.2$\cdot10^3$\\
$r_0$, \r{A}&2.1456&2.7$\cdot10^{-3}$&3.9$\cdot10^{-3}$&
4.0$\cdot10^{-3}$&$\hat r_0$&3.6$\cdot10^{-3}$\\
$E_0$, eV&4.42&0.23&0.34&0.35&$10$&0.21\\
\end{tabular}\end{minipage}\label{Errors}\end{table*}

We now turn our attention to the errors of fitting parameters. In ignoring the
correlations, the errors $\delta^{\rm(a)}p$ are rather small (see
Table~\ref{Errors}). However, we know that the parameters $r_0$ and $E_0$ are
highly correlated, and their real errors must be larger. In the traditional
XAFS-analysis two-dimensional contour maps have long been used \cite{Joyner1}
for estimates of the correlation score and the error bars. Notice, that to
do this requires, first, the definition and determination of the correct
statistical function $\chi^2$ (but not a proportionate to it), and, second, a
criterion to choose the critical value of $\chi^2$ (depending on the chosen
confidence level).

\begin{figure}[!t]\includegraphics*{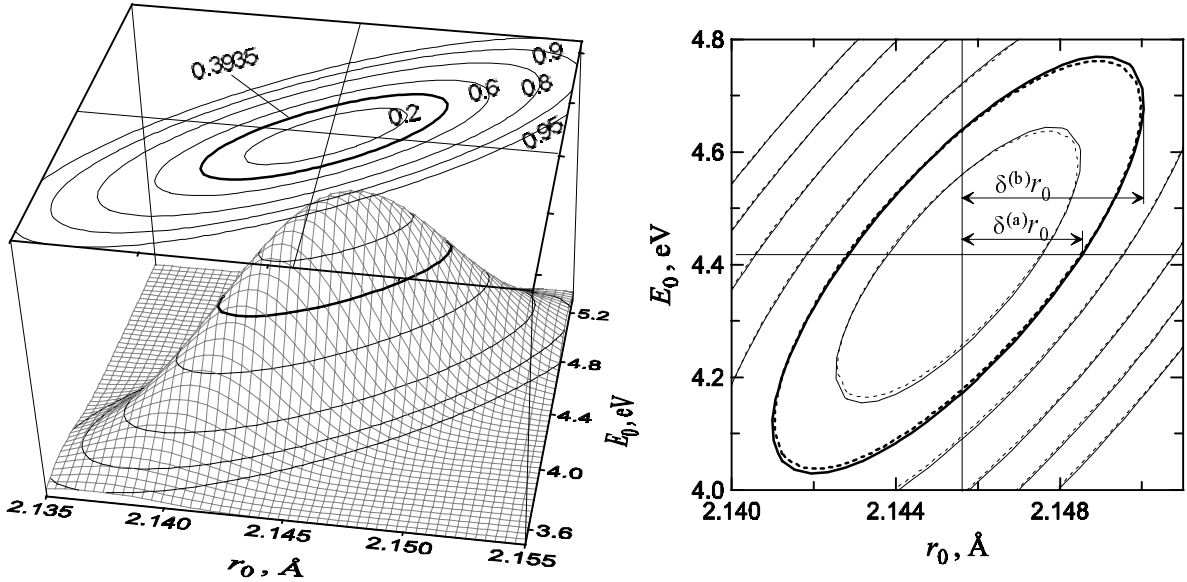}
\caption{The joint probability density function $P(r_0E_0|{\bf d})$ calculated
via the expansion \protect{(\ref{P4})} (solid lines) and using the exact
$\chi^2$ function (on the right, dashed lines). Also shown the graphical
interpretation of the mean-square deviations $\delta^{\rm(a)}r_0$ and
$\delta^{\rm(b)}r_0$ given by \protect{(\ref{P4_1_1})} and
\protect{(\ref{P4_3})}. The ellipse of the standard deviation is drawn by the
thick line.} \label{rE0}\end{figure}

For the most correlated pair, $r_0$ and $E_0$, find the joint probability
density function $P(r_0E_0|{\bf d})$ using the Hessian elements found at the
minimum of the $\chi^2$:
\begin{eqnarray}\label{S1}
P(r_0E_0|{\bf d})\propto\exp\Bigl(-\frac{1}{4}\tensor{\rm H}_{r_0r_0}(\Delta
r_0)^2-\frac{1}{2}\tensor{\rm H}_{r_0E_0}\Delta r_0\Delta E_0
-\frac{1}{4}\tensor{\rm H}_{E_0E_0}(\Delta E_0)^2\Bigr)
\end{eqnarray}
which is depicted in Fig.~\ref{rE0} as a surface graph and as a contour map.
The ellipses of the equal probability are described by:
\begin{eqnarray}\label{S2}
\tensor{\rm H}_{r_0r_0}(\Delta r_0)^2+2\tensor{\rm H}_{r_0E_0}\Delta
r_0\Delta E_0+\tensor{\rm H}_{E_0E_0}(\Delta E_0)^2=4\lambda.
\end{eqnarray}
In Fig.~\ref{rE0} they limit such areas that the probability for the random
vector ($r_0$,$E_0$) to find itself in them is equal to
$\ell=1-e^{-\lambda}=0.2$, 0.6, 0.8, 0.9 and 0.95. By the thick line is drawn
the ellipse corresponding to the standard deviation: $\lambda=1/2$ and
$\ell=1-e^{-1/2}\approx0.3935$. For this ellipse the point of intersection with
the line $\Delta E_0=0$ and the point of maximum distance from the line
$\Delta r_0=0$ give the standard mean-square deviations $\delta^{\rm(a)}r_0$ and
$\delta^{\rm(b)}r_0$ that coincide with the expressions (\ref{P4_1_1}) and
(\ref{P4_3}). To find the mean-square deviation $\delta^{\rm(b)}$ for an
arbitrary confidence level $\ell$, one should multiply the standard deviation
by $\sqrt{-2\ln(1-\ell)}$.

In Table~\ref{Errors} the errors in the column $\delta^{\rm(b)}p$ were found as
the largest errors among all those calculated from the pair correlations. For
the parameters $N$ and $a$ all pair correlations are weak, so their
$\delta^{\rm(a)}$ and $\delta^{\rm(b)}$ are hardly differ. For the parameters
$r_0$ and $E_0$ these mean-square deviations differ remarkable.

Finally, we put the question, how much is rightful the expansion (\ref{P4}) for
the likelihood function? In Fig.~\ref{rE0}, on the right, the dashed ellipses
of equal probability are found for the exact $\chi^2$ that was calculated by
the {\sc viper} program as well. Mainly, just-noticeable difference is caused
by the realization of the fitting algorithm or to be more precise, by the
values of the variations of the fitting parameters which determine the accuracy
of the minimum itself and the accuracy of the derivatives at the minimum. Of
course, this difference can be neglected.

\subsection{General case}

Often, a particular fitting parameter significantly correlates not with a one,
but with {\em several} other parameters (in our example this is not so, but,
for instance, the problem of approximation of the atomic-like background by
interpolation spline drawn through the varied knots \cite{Newville1,Ie} is that
very case). Now, the consideration of the two-dimensional probability density
functions is not correct no more, one should search for the {\em total}
joint posterior probability $P({\bf p}|{\bf d})$.

For that, first of all, one is to find the prior probability $P({\bf p})$.
Let we approximately know in advance the size $S_k$ of the variation range of
the parameter $p_k$. Then the prior probability can be expressed as:
\begin{equation}\label{P5}
P({\bf p}|\alpha)\propto\alpha^{N/2}\exp
\Bigl(-\frac{\alpha}{2}\sum_{k=1}^{N}\frac{\Delta p_k^2}{S_k^2}\Bigr),
\end{equation}
where the regularizer $\alpha$ specifies the relative weight of the prior
probability; at $\alpha=0$ there is no prior information, at $\alpha\to\infty$
the fitting procedure gives nothing and the posterior distribution coincides
with the prior one. In the expression (\ref{P5}) $\alpha$ appears as a known
value. Later, we apply the rules of probability theory to remove it from
the problem.

So, for the sought probability density functions we have:
\begin{eqnarray}\label{P6}
P(p_j|{\bf d},\alpha)&\propto&\int\cdots dp_{i\ne j}\cdots\alpha^{N/2}
\exp\Bigl(-\frac{1}{2}\sum_{k,l=1}^{N}g_{kl}\Delta p_k\Delta p_l\Bigr),
\end{eqnarray}
where
\begin{eqnarray}\label{P7}
g_{kl}=\frac{\alpha}{S_k^2}\delta_{kl}+\frac{\tensor{\rm H}_{kl}}{2}.
\end{eqnarray}
Since there is no integral over $p_j$, separate it from the other integration
variables:
\begin{eqnarray}\label{P8}
P(p_j|{\bf d},\alpha)&\propto&\alpha^{N/2}
\exp\Bigl(-\frac{1}{2}g_{jj}\Delta p_j^2\Bigr)
\int\cdots dp_{i\ne j}\cdots
\exp\Bigl(-\frac{1}{2}\mathop{{\sum}\smash{^j}}_{k,l=1}^{N}
g_{kl}\Delta p_k\Delta p_l-
\Delta p_j\mathop{{\sum}\smash{^j}}_{k=1}^{N}g_{kj}\Delta p_k\Bigr),
\end{eqnarray}
Here, the symbol $j$ near the summation signs denotes the absence of $j$-th
item. Further, find the eigenvalues $\lambda_i$ and corresponding eigenvectors
${\bf e}_i$ of the matrix $g_{kl}$ in which the $j$-th row and column are
deleted, and change the variables:
\begin{equation}\label{P9}
b_i=\sqrt{\lambda_i}\mathop{{\sum}\smash{^j}}_{k=1}^{N}\Delta p_ke_{ik},\qquad
\Delta p_k=\mathop{{\sum}\smash{^j}}_{i=1}^{N}\frac{b_ie_{ik}}{\sqrt{\lambda_i}}
\qquad (i,k\ne j).
\end{equation}
Using the properties of eigenvectors:
\begin{equation}\label{P10}
\mathop{{\sum}\smash{^j}}_{k=1}^{N}g_{lk}e_{ik}=\lambda_ie_{il},\qquad
\mathop{{\sum}\smash{^j}}_{k=1}^{N}e_{lk}e_{ik}=\delta_{li}\qquad (l,i\ne j),
\end{equation}
one obtains:
\begin{eqnarray}\label{P11}
P(p_j|{\bf d},\alpha)&\propto&\alpha^{N/2}
\exp\Bigl(-\frac{1}{2}[g_{jj}-{\bf u}^2]\Delta p_j^2\Bigr)
\int\cdots db_{l\ne j}\cdots
\exp\Bigl(-\frac{1}{2}\mathop{{\sum}\smash{^j}}_{i=1}^{N}
[b_i+u_i\Delta p_j]^2\Bigr)\nonumber\\&\propto&\alpha^{N/2}
\exp\Bigl(-\frac{1}{2}[g_{jj}-{\bf u}^2]\Delta p_j^2\Bigr),
\end{eqnarray}
where new quantities were introduced:
\begin{eqnarray}\label{P12}
u_i=\frac{1}{\sqrt{\lambda_i}}\mathop{{\sum}\smash{^j}}_{k=1}^{N}g_{kj}e_{ik},\qquad
{\bf u}^2=\mathop{{\sum}\smash{^j}}_{i=1}^{N}u_i^2.
\end{eqnarray}

Thus, we have found the explicit expression for the posterior distribution of
an arbitrary fitting parameter. This is a Gaussian distribution with the mean
$\bar p_j=\hat p_j$ and the standard deviation
\begin{eqnarray}\label{P13}
\delta^{\rm(c)}p_j=(g_{jj}-{\bf u}^2)^{-1/2}.
\end{eqnarray}

The formulas (\ref{P11})--(\ref{P13}) require to find the eigenvalues and
eigenvectors for the matrix of rank $N-1$ for each parameter. Those formulas
have merely a methodological value: the explicit expressions for posterior
probabilities enables one to find the average of {\em arbitrary} function of
$p_j$. However, the standard deviations could be calculated significantly
easier, having found the eigenvalues and eigenvectors for the matrix of
rank $N$ one time.
\begin{eqnarray}\label{P14}
(\delta^{\rm(c)}p_j)^2=
\frac{\int\Delta p_j^2P(p_j|{\bf d},\alpha)dp_j}{\int P(p_j|{\bf d},\alpha)dp_j}=
\frac{\int\Delta p_j^2\exp\Bigl(-\frac{1}{2}
\sum_{k,l=1}^{N}g_{kl}\Delta p_k\Delta p_l\Bigr)d{\bf p}}
{\int\exp\Bigl(-\frac{1}{2}\sum_{k,l=1}^{N}g_{kl}\Delta p_k\Delta
p_l\Bigr)d{\bf p}}.
\end{eqnarray}
Analogously to what was done above, performing the diagonalization of $g_{kl}$,
one obtains:
\begin{eqnarray}\label{P15}
(\delta^{\rm(c)}p_j)^2=\frac
{\int d{\bf b}\Bigl(\sum_{i=1}^{N}b_ie_{ij}/{\sqrt{\lambda_i}}\Bigr)^2
\exp\Bigl(-\frac{1}{2}\sum_{i=1}^{N}b_i^2\Bigr)}
{\int d{\bf b}\exp\Bigl(-\frac{1}{2}\sum_{i=1}^{N}b_i^2\Bigr)}
=\sum_{i=1}^{N}\frac{e_{ij}^2}{\lambda_i},
\end{eqnarray}
where the eigenvalues ($\lambda_i$) and eigenvectors (${\bf e}_i$) correspond
to the full matrix $g_{kl}$.

One can give another interpretation of the $\delta^{\rm(c)}p$-finding process.
It is easy to verify that $\tensor{\rm H}/2$ and the covariance matrix $C$ of
the vector ${\bf p}$ are mutually inverse. Therefore
\begin{eqnarray}\label{P15_1}
(\delta^{\rm(c)}p_j)^2=C_{jj}=2(\tensor{\rm H}^{-1})_{jj},
\end{eqnarray}
and the variate $({\bf p-\hat p})^T\cdot C^{-1}\cdot({\bf p-\hat p})=
\frac{1}{2}({\bf p-\hat p})^T\cdot\tensor{\rm H}\cdot({\bf p-\hat p})$ is
$\chi^2$-distributed with $N$ degrees of freedom if ${\bf p}$ is the
$N$-dimensional normally distributed vector (by Eq.~(\ref{P11}) this condition
is met). The ellipsoid that determines the standard deviation is:
\begin{eqnarray}\label{P15_2}
({\bf p-\hat p})^T\cdot\tensor{\rm H}\cdot({\bf p-\hat p})=N.
\end{eqnarray}
For an arbitrary confidence level $\ell$, on the r.h.s. would be
$(\chi^2_N)_\ell$, the critical value of the $\chi^2$-distribution with $N$
degrees of freedom. The error $\delta^{\rm(c)}p_k$ is equal to the half the
ellipsoid size along the $k$-th axis.

In our example fitting, the errors found in the absence of any prior
information ($\alpha=0$) from the formula (\ref{P15}) are listed in
Table~\ref{Errors} in the column $\delta^{\rm(c)}p$. Due to every one parameter
correlates at the most with one other parameter, all $\delta^{\rm(c)}p$ are
practically coincide with $\delta^{\rm(b)}p$. Generally, this may be not so.

\begin{figure}[!b]
\begin{minipage}[c]{0.5\hsize}\includegraphics*{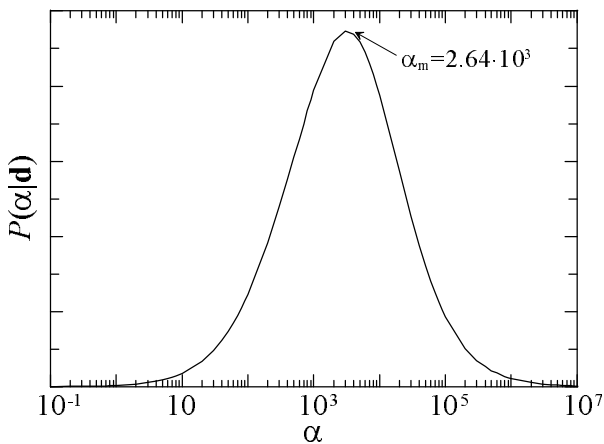}\end{minipage}
\begin{minipage}[c]{0.45\hsize}\caption{The posterior distribution for the
regularizer $\alpha$ found from Eq.~(\protect{\ref{P17}}).}
\label{alpha}\end{minipage}\end{figure}
Finally, let us find the most probable value of $\alpha$. Its posterior
distribution is given by:
\begin{eqnarray}\label{P16}
P(\alpha|{\bf d})=\int d{\bf p}P(\alpha,{\bf p}|{\bf d})=
\int d{\bf p}P(\alpha)P({\bf p}|\alpha,{\bf d}).
\end{eqnarray}
Using a Jeffreys prior $P(\alpha)=1/\alpha$ \cite{Jeffreys1}, one obtains
for the posterior distribution:
\begin{eqnarray}\label{P17}
P(\alpha|{\bf d})&\propto&\int d{\bf p}\alpha^{N/2-1}
\exp\Bigl(-\frac{1}{2}\sum_{k,l=1}^{N}g_{kl}\Delta p_k\Delta p_l\Bigr)
\propto (\lambda_1\cdots\lambda_N)^{-1/2}\alpha^{N/2-1}.
\end{eqnarray}

In our example we have set the variation range of the parameter $p_k$ to be
equal to $S_k=\pm\hat p_k$ (this means that $p_k\in [0,2\hat p_k]$) for all
parameters except for $E_0$; since it varies near zero, we have chosen
$S_{E_0}=\pm10$\,eV. For the mentioned variation ranges, the distribution
$P(\alpha|{\bf d})$ has its mode at $\alpha=2.64\cdot10^3$ (see
Fig.~\ref{alpha}). The bayesian errors found for this regularizer are listed in
the column $\delta^{\rm(d)}p$ of Table~\ref{Errors}. As a result, we have got
the mean-square errors that for some parameters are significantly lower than
even $\delta^{\rm(a)}p$. There is nothing surprising in that: any
additional information narrows the posterior distribution. If we would choose
$S_k$ to be less, $\delta^{\rm(d)}p_k$ would be yet lower. For instance, XAFS
is quite accurate in distance determination, and for many cases one can assume
distances to be known within $\pm0.2$\,\r{A}. In our case this leads to
$\delta^{\rm(d)}r_0=3.4\cdot10^{-3}$\,\r{A}.

\subsection{Important note}

Having obtained the expressions (\ref{P4_1_1}), (\ref{P4_3}) and (\ref{P15})
for the errors of fitting parameters, we are able now to draw an important
conclusion. If in the definition (\ref{A4}) one substitutes for $\varepsilon_i$
the values that are smaller by a factor of $\beta$ than the real ones, the
$\chi^2$ and its Hessian's elements are exaggerated by a factor of $\beta^2$,
and from (\ref{P4_1_1}), (\ref{P4_3}) and (\ref{P15}) follows that the errors
of fitting parameters are understated by a factor of $\beta$!

In the preceding paper \cite{Ie} it was shown that the errors of the
atomic-like absorption construction are essentially larger than the experimental
noise, and therefore it is the former that should determine the $\varepsilon_i$
values. However, these values are traditionally assumed to be equal to the
noise, or one uses unjustified approximations for them, also understated
(like $1/\varepsilon_i^2=k^w$ \cite{Filipponi1}). It is here where we see the
main source of the groundless optimistic errors.

\section{Statistical tests in fitting problems}
\subsection{$\chi^2$-test}
Introducing the statistical function $\chi^2$, we assumed that it follows the
$\chi^2$ distribution with $\nu=M-N$ degrees of freedom. However for this would
be really so, one should achieve a sufficient fitting quality. This
``sufficient quality'' could be defined as such that the variate (\ref{A4})
obeys the $\chi^2$ distribution law, that is this variate does not fall within
the tail of this distribution. Strictly speaking, the following condition must
be met:
\begin{eqnarray}\label{T1}
\chi^2<(\chi^2_\nu)_\ell,
\end{eqnarray}
where the critical value $(\chi^2_\nu)_\ell$ for the specified significance
level $\ell$ may be calculated exactly (for even $\nu$) or approximately (for
odd $\nu$) using the known formulas \cite{Handbook}.

Notice, that the choice of the true $\varepsilon_i$ here also plays a cardinal
role. However, it is important here that one would not use the {\em
overestimated} values which facilitate to meet the requirement (\ref{T1}). As
we have shown in \cite{Ie}, one could obtain the overestimated $\varepsilon_i$,
having assumed the Poisson destribution law for the detectors counts when the
actual association between the probability of a single count event and the
radiation intensity is unknown.

Thus, the exaggerated values $\varepsilon_i$ tell about a quality fitting, but
give the large errors of fitting parameters. The understated $\varepsilon_i$
lead to the would-be small errors, but make difficult to pass the $\chi^2$-test
(i. e. to meet the condition (\ref{T1})). We are aware of many works the
authors of which do not describe explicitly the evaluation process for the
errors of XAFS-function extraction and do not report their explicit values.
However, by implication it is seen that $\varepsilon_i$ were chosen (not
calculated!) as low as possible to scarcely (with $\ell=0.9-0.95$) pass the
$\chi^2$-test; as a result, very impressive errors of the structural parameters
were obtained. In such approach no wander that the difference of 0.01\,\r{A}
between the diffraction data and the XAFS-result that was found within
0.002\,\r{A} was attributed to the ``suggested presence of a small systematic
error'' \cite{Filipponi1}.

\subsection{$F$-test}
Let there is a possibility to choose between two physical models depending on
different numbers of parameters $N_1$ and $N_2$ ($N_2>N_1$). Which one of them
is more statistically important? For instance one wish to decide whether a
single coordination sphere is split into two.

\begin{figure}[!t]
\begin{minipage}[c]{0.5\hsize}\includegraphics*{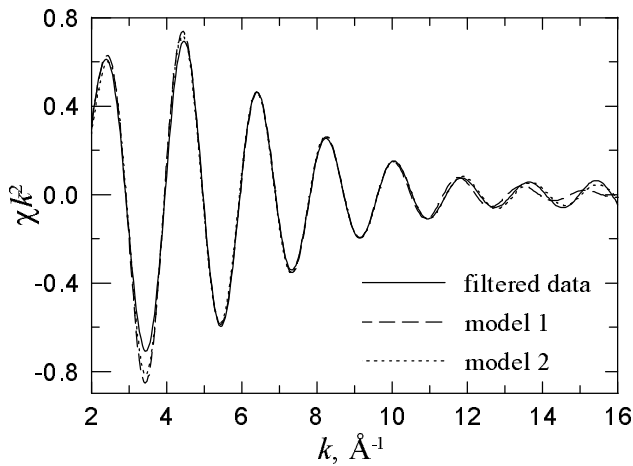}\end{minipage}
\begin{minipage}[c]{0.45\hsize}\caption{On the choice between two different
models on statistical grounds. Cited from Ref.~\protect\cite{MenushenP}.}
\label{tests}\end{minipage}\end{figure}
Let for the two models the functions $\chi^2_1$ and $\chi^2_2$ obey the
$\chi^2$-distribution law with $\nu_1=N_{\rm ind}-N_1$ and
$\nu_2=N_{\rm ind}-N_2$ degrees of freedom, correspondingly. From the linear
regression problem (near the minimum of $\chi^2$, the likelihood function is
expressed by (\ref{P4}) and is identical in form to that of the linear
regression problem) it is known that the value
\begin{eqnarray}\label{F1}
f=\frac{(\chi^2_1-\chi^2_2)/(\nu_1-\nu_2)}{\chi^2_2/\nu_2}
\end{eqnarray}
obeys the Fisher's $F$-distribution law with $(\nu_1-\nu_2,\nu_2)$ degrees of
freedom if exactly $r=\nu_1-\nu_2$ parameters in the second model are linearly
dependent, that is if exist the $r\times N_2$ matrix $C$ of rank $r$ and the
vector $\bf c$ of the dimension $r$ such that $C{\bf p}={\bf c}$. In order for
the linear restrictions on the second model parameters to be absent, the value
$f$ should {\em not} follow the $F$-distribution, that is it should be greater
than the critical value $(F_{\nu_1-\nu_2,\nu_2})_\ell$ for the specified
significance level $\ell$:
\begin{eqnarray}\label{F2}
f>(F_{\nu_1-\nu_2,\nu_2})_\ell
\end{eqnarray}
or
\begin{eqnarray}\label{F3}
\chi^2_2<\chi^2_1\left((F_{\nu_1-\nu_2,\nu_2})_\ell
\frac{\nu_1-\nu_2}{\nu_2}+1\right)^{-1}.
\end{eqnarray}
Notice, that the expression (\ref{F3}) means the absence of exactly $r$ linear
restrictions on the second model parameters. Even if (\ref{F3}) is realized,
the less number of linear dependencies are possible. If, for instance,
the splitting of a single coordination sphere into two does not contradict to
the $F$-test (\ref{F3}), some of the parameters of these two spheres may be
dependent, but not all. This justifies the introduction of a new sphere into
the model XAFS function.

Thus, having specified the significance level $\ell$, one can answer the
question ``what decrease of $\chi^2$ must be achieved to increase the number of
parameters from $N_1$ to $N_2$?'' or, inside out, ``what is the probability that
the model 2 is better than the model 1 at specified $(N_1,\chi^2_1)$ and
$(N_2,\chi^2_2)$?''

Notice, that since in the definition for $f$ the ratio $\chi^2_1/\chi^2_2$
appears, the actual values of $\varepsilon_i$ become not important for the
$F$-test (only if they all are taken equal to a single value).

Consider an example of the statistical tests in the fitting problem. In
Fig.~\ref{tests} are shown the experimental curve with $N_{\rm ind}=11.8$ and
two model curves with $N_1=4$ and $N_2=7$. The underlying physical models were
described in Ref.~\cite{MenushenP}; here only the number of parameters is of
importance. Let us apply the statistical tests. Through the fitting procedure
for the model 1 we have:  $\nu_1=11-4=7$,
$\chi^2_1=16.8>14.1=(\chi^2_7)_{0.95}$, for the model 2:  $\nu_2=11-7=4$,
$\chi^2_1=5.3<9.5=(\chi^2_4)_{0.95}$. That is the first model does not pass the
$\chi^2$-test. Further, $f=2.89=(F_{3, 4})_{0.84}$, from where with the
probability of 84\% we can assert that the model 2 is better than the model 1.

In the XAFS analysis the $F$-test has long been in use \cite{Joyner1}. However,
the words substantiating the test are often wrong. The authors of Refs.
\cite{Filipponi1,Michalowicz1}, for example, even claimed that the value $f$
(\ref{F1}) {\em must} follow the $F$-distribution, although then in
Ref.~\cite{Michalowicz1} there appears a really correct inequality (\ref{F3}).

\section{Conclusion}
The solution of the main task of the XAFS spectroscopy, determination of the
structural parameters, becomes worthless if the confidence in this solution is
unknown. Here we mean not only the confidence in the obtained fitting parameters
that is their mean-square deviations, but also the credence to the very methods
of the error analysis. It is excessive optimistic errors evaluations lead to
the suspicious attitude to the XAFS results.

To improve the situation could the development of the reliable and well-grounded
techniques that do not allow one to treat the data in an arbitrary way.
First of all, this is a technique for determination of the real errors of the
atomic-like absorption construction. Second, we regard as necessary to
standardize the method for the correct taking into account of {\em all} pair
correlation between fitting parameters. And third, (we have not raised this
question here) programs for scattering phase and amplitude calculations should
report on the confidence limits for the calculated values, that is report how
sensitive the calculated values are to the choice of the parameters of
scattering potentials.


\begin{thebibliography}{10}

\bibitem{Young1}
N.~A. Young, A.~J. Dent, {\em Open Letter to the XAFS Comunity. Maintaining and
  improving the quality of published XAFS data: a view from the UK XAFS user
  group}. J. Synchrotron Rad. {\bf 6}, 799  (1999), (Proc. of Int. Conf. XAFS
  X).

\bibitem{Ie}
K.~V. Klementev, {\em XAFS analysis. I. Extracting the fine structure from the
  absorption spectra}. The preceding article ,   (2000).

\bibitem{Catalog}
Catalog of XAFS Analysis Programs, {\tt
  http://ixs.csrri.iit.edu/catalog/XAFS\verb"_"Programs }.

\bibitem{Stern1}
E.~A. Stern, {\em Number of relevant independent points in x-ray-absorption
  fine-structure spectra}. Phys. Rev. B {\bf 48}(13), 9825--9827  (1993).

\bibitem{FEFF}
J.~J. Rehr, J.~Mustre de~Leon, S.~I. Zabinsky, R.~C. Albers, {\em Theoretical
  X-ray Absorption Fine Structure Standards}. J.~Am. Chem. Soc. {\bf 113},
  5135--5140  (1991).

\bibitem{VIPER}
K.~V. Klementev, {\em VIPER for Windows (Visual Processing in EXAFS
  Researches)}, \\freeware, {\tt
  http://www.crosswinds.net/\symbol{126}klmn/viper.html }.

\bibitem{Joyner1}
R.~W. Joyner, K.~J. Martin, P. Meehan, {\em Some applications of statistical
  tests in analysis of EXAFS and SEXAFS data}. J.~Phys.~C: Solid State Phys.
  {\bf 20}, 4005--4012  (1987).

\bibitem{Newville1}
M. Newville, P. {L\=\i vi\c{n}\v{s}}, Y. Yacoby, J.~J. Rehr, E.~A. Stern, {\em
  Near-edge x-ray-absorption fine structure of {Pb}: {A}~comparison of theory
  and experiment}. Phys. Rev. B {\bf 47}(21), 14126--14131  (1993).

\bibitem{Jeffreys1}
H. Jeffreys, Theory of Probability (Oxford University Press, London, 1939),
  later editions: 1948, 1961, 1983.

\bibitem{Filipponi1}
A. Filipponi, A.~Di Chicco, {\em X-ray-absorption spectroscopy and $n$-body
  distribution functions in condensed matter. II. Data analysis and
  applications}. Phys. Rev. B {\bf 52}, 15135--15149  (1995).

\bibitem{Handbook}
Handbook of mathematical functions with formulas, graphs and mathematical
  tables, edited by M. Abramowitz, I. Stegun (Applied mathematical series, 55,
  National bureau of standards, 1964).

\bibitem{MenushenP}
A.~P. Menushenkov, K.~V. Klementev, {\em EXAFS indication of double-well
  potential for oxygen vibration in Ba$_{1-x}$K$_x$BiO$_3$}. J.~Phys.: Condens.
  Matter {\bf 12},   (2000), (accepted).

\bibitem{Michalowicz1}
A. Michalowicz, K. Provost, S. Laruelle, A. Mimouni, {\em F-test in EXAFS
  fitting of structural models}. J. Synchrotron Rad. {\bf 6}, 233--235  (1999),
  (Proc. of Int. Conf. XAFS X).

\end{thebibliography}

\end{document}